# First observation of electric-quadrupole infrared transitions in water vapour

Alain Campargue [a] [*], Samir Kassi [a], Andrey Yachmenev [b,c],
Aleksandra A. Kyuberis [d], Jochen Küpper [b,c,e], Sergei N. Yurchenko [f]

[a] *Univ. Grenoble Alpes, CNRS, LIPhy, 38000 Grenoble, France*
[b] *Center for Free-Electron Laser Science, Deutsches Elektronen-Synchrotron DESY, Notkestraße 85, D-22607 Hamburg, Germany*
[c] *Center for Ultrafast Imaging, Universität Hamburg, Luruper Chaussee 149, 22761 Hamburg, Germany*
[d] *Institute of Applied Physics, Russian Academy of Sciences, Ulyanov Street 46, Nizhny Novgorod, Russia 603950*
[e] *Department of Physics, Universität Hamburg, Luruper Chaussee 149, 22761 Hamburg, Germany*
[f] *Department of Physics and Astronomy, University College London, London, WC1E 6BT, UK*

Number of Pages:    9
Number of Figures:  2
Number of Tables:   2

Corresponding author.
E-mail address: alain.campargue@univ-grenoble-alpes.fr
Tel.: 33 4 76 51 43 19   Fax. 33 4 76 63 54 95

*Molecular absorption of infrared radiation is generally due to ro-vibrational electric-dipole transitions. Electric-quadrupole transitions may still occur, but they are typically a million times weaker than electric-dipole transitions, rendering their observation extremely challenging. In polyatomic or polar diatomic molecules, ro-vibrational quadrupole transitions have never been observed. Here, we report the first direct detection of quadrupole transitions in water vapor. The detected quadrupole lines have intensity largely above the standard dipole intensity cut-off of spectroscopic databases and thus are important for accurate atmospheric and astronomical remote sensing.*

Spectroscopic techniques in the IR domain have advanced significantly in recent years providing exciting opportunities for new generation of accurate and sensitive measurements. Quadrupole electronic and ro-vibrational transitions have been detected in a number of atoms [1,2] and homonuclear diatomic molecules [3,4,5], respectively, for states that cannot undergo electric dipole transitions. When compared with dipole transitions, quadrupole transition rates are typically smaller by a factor ranging from $10^8$ in the IR to $10^5$ in the optical spectral domain. Such extremely long mean lifetimes against spontaneous emission of quadrupole excitations make them ideal candidates for next-generation atomic clocks [6], high-precision tests of molecular physics [7,8,9], and quantum information processing [6,10]. Electric-quadrupole transitions are used for remote sensing of important diatomic molecules, such as hydrogen, oxygen, and nitrogen, in spectra of Earth's atmosphere and other environments [4,11,12].

In most heteronuclear diatomic and polyatomic molecules the ro-vibrational states can be excited both *via* dipole and quadrupole transitions. The dipole and quadrupole moment operators have distinct selection rules. The dipole moment connects states with different parity, while for the quadrupole transitions, the parity is conserved. Furthermore, states with rotational quantum numbers $J$ differing by ±2 can only be connected by a quadrupole transition. This difference in selection rules permits, in principle, the observation of well-isolated quadrupole lines. Nevertheless, in practice the detection of quadrupole lines is made difficult by the fact that, even if transition frequencies differ, the quadrupole lines are drowned in the line profile of the much stronger dipole lines.

To the best of our knowledge, quadrupole transitions have never been detected in polyatomic molecules until now, although the quadrupole moment of some polyatomic molecules ($CO_2$, $NO_2$, OCS) have been measured using electric-field-gradient-induced birefringence method [13]. Nowadays, spectroscopic line lists of the major greenhouse gas absorbers such as water vapour and carbon dioxide, are limited to dipole transitions [14]. A highly-accurate and complete characterization of ro-vibrational spectra of these molecules is needed for the modeling and understanding of Earth's atmosphere, climate, and many remote sensing experiments [15]. At present, adequate models of the water vapour absorption in the Earth's atmosphere incorporate many weak dipole lines including hot band transitions, minor water isotopologues contribution, and weak absorption continua [16]. Obviously, failure to include weak quadrupole transitions to the total opacities of atmospherically important species may contribute to systematic errors in remote sensing retrievals with a corresponding impact on atmospheric simulations.

Here, we report the first observation of electric-quadrupole transitions in water vapour. This is actually the first time that quadrupole lines of a polyatomic molecule have been detected in an IR spectrum. This observation was made possible using a continuous wave diode laser cavity ring down spectroscopic (CRDS) technique to measure the water absorption spectrum near 1.3 μm. The detection and assignment of the measured spectral lines relied on high-accuracy predictions for both positions and intensities of the quadrupole $H_2^{16}O$ transitions. It is important to stress that the detected quadrupole lines have intensity largely above the recommended dipole intensity cut-off of the standard spectroscopic databases and should thus be incorporated.

CRDS is a high-sensitivity absorption technique which performs like a classical absorption spectroscopy that would have a multi-kilometric absorption path-length. The method consists in measuring the power decay rate (ring down) of a laser light trapped into a passive optical cavity, made of high-reflectivity semi-transparent mirrors [17]. The presence of an absorbing gas in the cavity shortens the ring down time. From the variation of the decay rate with the laser frequency, one gets the absorption spectrum. Because of the extreme reflectivity of the mirrors (typically $R\sim99.997$ %) sensitivities equivalent to several hundred kilometers in a classical absorption approach are routinely achieved and extremely weak absorptions are detected. A record absorption sensitivity, $\alpha_{min} \approx 5\times10^{-13}$ cm$^{-1}$, corresponding to a 2% light attenuation for the Earth-Moon distance, was achieved by CRDS [18], allowing for the detection of extremely weak quadrupole lines of $D_2$ and $N_2$ [19]. In the last two decades, the CRDS technique was extensively used to characterize weak dipole transitions of water vapour in the near-IR spectral domain [20]. Tens of thousands of new dipole transition lines were detected and assigned to various water isotopologues present in natural isotopic abundance. In the 1.6 and 1.25 μm transparency windows, lines of HD$^{17}$O with a natural relative abundance of $1.158\times10^{-7}$ were detected [21] and dipole lines with intensities as weak as $1\times10^{-29}$ cm/molecule were measured [22].

At room temperature and typical sample cell pressures in the Torr range, most of quadrupole transitions of water molecule are masked by much stronger dipole transitions. In order to detect them in the CRDS spectrum, a highly accurate theoretical prediction of the quadrupole ro-vibrational spectrum is essential.

Here, the calculations employed a state-of-the-art variational approach based on high-level *ab initio* electric dipole and quadrupole moment surfaces of $H_2O$ [23]. The robust variational approach TROVE [24,25] was used to solve the ro-vibrational Schrödinger equation for the motion of nuclei, based on exact kinetic energy operator [26,27] and recently reported accurate potential energy surface of $H_2O$ [26] (see supplementary material [30]). The electric-quadrupole moment surface was computed *ab initio* at the CCSD(T)/aug-cc-pVQZ level of theory in the frozen-core approximation using the CFOUR program package [27]. The quadrupole intensities and Einstein *A* coefficients were calculated using the generalized approach RichMol for molecular dynamics in the presence of external electric fields [28]. The basis-set convergence of the ro-vibrational line positions was carefully examined. The remaining errors on the order of 0.01-0.6 cm$^{-1}$ were still too significant to unambiguously identify quadrupole lines in the CRDS spectrum. To improve the accuracy, the computed ro-vibrational line positions were adjusted according to empirical values of the lower and upper state energy levels of $H_2^{16}O$. These energies were carefully derived from the analysis of high-resolution spectroscopic data from more than a hundred experimental sources [29]. The average uncertainties on the order $10^{-3}$ cm$^{-1}$ in the resulting line centers allowed to unambiguously identify quadrupole lines. Indeed, in the region of interest, with the sensitivity of the CRDS spectra under analysis, the spectral congestion is very high, due to dipole lines of $H_2^{16}O$, water minor isotopologues, as well as impurities such as $CO_2$, $NH_3$ or $CH_4$ present at ppm relative concentrations.

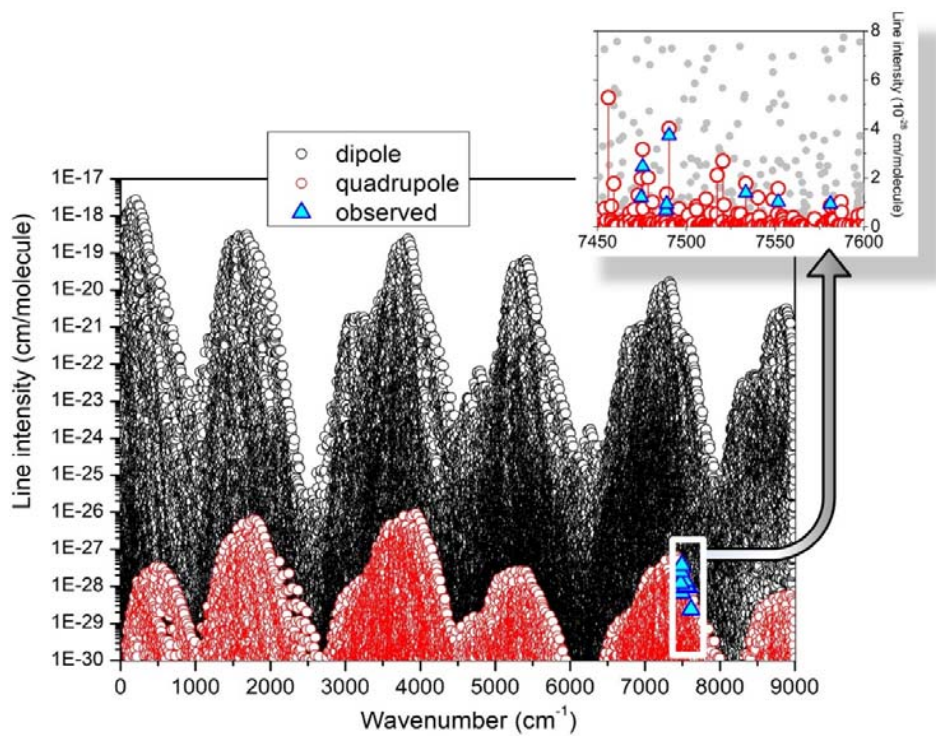

**Fig. 1** Overview of the absorption line list of $H_2^{16}O$ at 296 K. The calculated electric-quadrupole spectrum [30] (red circles) is superimposed to the calculated electric-dipole spectrum (black circles). The CRDS measured quadrupole transitions are shown with blue triangles. The insert with intensities in linear scale shows a zoom of the 7450-7600 cm$^{-1}$ region.

The overview of the calculated quadrupole line list of $H_2^{16}O$ [30] is displayed shown in **Fig. 1** superimposed to the dipole line list computed using an *ab initio* dipole moment surface [31]. Both lists include all ro-vibrational transitions of $H_2^{16}O$ with $J \leq 30$ and upper state energies below 10000 cm$^{-1}$ with respect to the zero-point level. The general appearance of the quadrupole and dipole line lists is similar; the spectra consist of a succession of vibrational bands whose intensity decreases with the frequency. The maximum intensity values of quadrupole lines are typically seven orders of magnitude smaller than the dipole lines. Note that the strongest quadrupole lines near 4000 cm$^{-1}$ have an intensity on the order of 10$^{-26}$ cm/molecule while the intensity cut off the water dipole transitions included in the HITRAN database [14] is three orders of magnitude smaller (10$^{-29}$ cm/molecule). Interestingly, the largest quadrupole-to-dipole intensity ratios of about 10$^{-4}$ are predicted for the water transparency window at 2500 cm$^{-1}$, dominated by transitions of the bending $v_2$ vibrational band. The complete quadrupole line list with transition frequencies up to 10000 cm$^{-1}$ and 10$^{-40}$ cm/molecule intensity cut-off is provided as Supplemental Material [30].

The CRDS recordings of natural water vapour were performed in the frequency range between 7408 and 7619 cm$^{-1}$ at a pressure limited to 1.0 Torr and temperature maintained at 296 K [32]. The noise equivalent absorption of the recordings, estimated as a root-mean-square deviation of the spectrum base line, is $\alpha_{min} \approx 5 \times 10^{-11}$ cm$^{-1}$. In terms of line intensity, this converts to a detectivity threshold on the order of few 10$^{-29}$ cm/molecule, well below the calculated intensities of the strongest quadrupole lines in the region (up to $6 \times 10^{-28}$ cm/molecule).

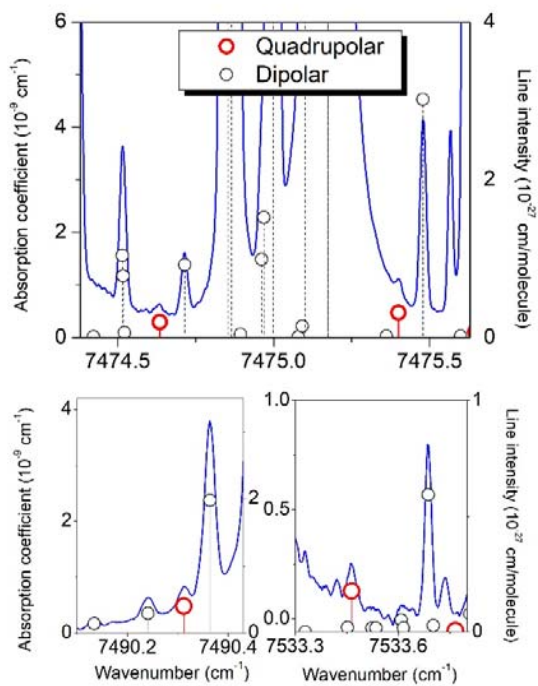

**Fig. 2** Detection of electric-quadrupole lines of H$_2$$^{16}$O in the spectrum of water vapour recorded by CRDS at 1.0 Torr. The quadrupole spectrum of H$_2$$^{16}$O (red stars) calculated in this work is superimposed to the dipole stick spectrum of water in natural isotopic abundance [14] (blue circles). The four detected quadrupole lines correspond to $\Delta J = 2$ transitions of the $v_1+v_3$ band.

As illustrated on **Fig. 2**, the remarkable position and intensity coincidences of the calculated quadrupole lines to the recorded spectra leaves no doubt that H$_2$$^{16}$O quadrupole lines are detected. Among the fifty quadrupole lines predicted with sufficient intensity in the considered region, nine could be detected (blue triangles in **Fig. 1**), the others being hidden by stronger dipole lines. As a result, **Table 1** lists the detected quadrupole lines with their measured [32] positions and intensities. The rovibrational assignments included in the table indicate that the detected lines belong to the $v_1+v_3$ band and correspond to $\Delta J= 2$ transitions except one line corresponding to $\Delta J= 1$. Experimental intensity values show a reasonable agreement with the calculated values [30] (see insert in **Fig. 1**). Excluding the two weakest lines with large experimental uncertainties, the average measured/computed intensity ratio is 0.80(17).

| Position (cm$^{-1}$) | | Int. (cm/molecule) | | $J\ K_a\ K_c$ | | | | | |
| --- | --- | --- | --- | --- | --- | --- | --- | --- | --- |
| Meas. | Calc. | Meas. | Calc. | Upper level | | | Lower level | | |
| 7474.6325 | 7474.6354 | 1.21E-28 | 1.98 E-28 | 6 | 1 | 5 | 4 | 2 | 3 |
| 7475.4020 | 7475.4005 | 2.46E-28 | 3.16 E-28 | 5 | 2 | 4 | 3 | 1 | 2 |
| 7488.5747 | 7488.5785 | 6.72E-29 | 3.23 E-29 | 6 | 6 | 1 | 5 | 5 | 0 |
| 7488.9183 | 7488.9215 | 9.34E-29 | 1.34 E-28 | 7 | 0 | 7 | 5 | 1 | 5 |
| 7490.3117 | 7490.3120 | 3.74E-28 | 4.02 E-28 | 7 | 1 | 7 | 5 | 0 | 5 |
| 7533.4649 | 7533.4642 | 1.39E-28 | 1.72 E-28 | 7 | 2 | 6 | 5 | 1 | 4 |
| 7551.7653 | 7551.7639 | 1.02E-28 | 1.56 E-28 | 9 | 1 | 9 | 7 | 0 | 7 |
| 7581.1247 | 7581.1172 | 9.32E-29 | 8.53 E-29 | 10 | 0 | 10 | 8 | 1 | 8 |
| 7613.8512 | 7613.8478 | 2.37E-29 | 1.10 E-29 | 10 | 2 | 9 | 8 | 1 | 7 |

**Table 1**
Assignment, transition frequencies, and intensities of the electric-quadrupole lines of the $\nu_1+\nu_3$ band of $H_2^{16}O$ measured near 1.3 µm.

This first detection of quadrupole transitions in the water vapour spectrum opens new perspectives for theoretical and experimental absorption spectroscopy of polyatomic molecules with expected impact in atmospheric and astronomical sciences. Despite being very weak, quadrupole absorption lines of water vapour are clearly above the background formed by the dipole spectrum and the water continuum and therefore must be included into modern atmospheric databases such as HITRAN [14] or ExoMol [33]. Water vapour, being one of the major interfering species for the trace gas analysis in air, missing quadrupole transitions may skew the optical measurement of the targeted species. Similar biases may be encountered for metrological spectroscopic measurements of line intensities or line shapes (*e.g.* Doppler-broadening thermometry [34,35]).

In the case of water vapour the envelopes of the strong quadrupole and dipole vibrational bands mostly coincide as a result of molecular symmetry and corresponding selection rules. This will not be the case in other important atmospheric species such as $CO_2$ where, due to the linear structure of the molecule, different selection rules apply to quadrupole and dipole bands. Relatively strong quadrupole bands may be located in spectral regions where dipole absorption is weak, the so-called transparency windows. In these cases, the importance of quadrupole transitions will even be more apparent.

This study paves the way to systematic computation of quadrupole spectra for all standard atmospheric molecules, *e.g.*, $CO_2$, $N_2O$, $CH_4$, HCN, CO, HF. Owing to the weak character of the quadrupole spectra, systematic calculations across wide spectral range are thus required to evaluate the significance of quadrupole transitions for atmospheric, astrophysical, and industrial applications. After validation tests by high-sensitivity measurements, quadrupole transitions should be systematically incorporated into the most currently used spectroscopic databases.

SY acknowledges support from the UK Science and Technology Research Council (STFC) No. ST/R000476/1. A substantial part of the calculations was performed using high performance computing facilities provided by DiRAC for particle physics, astrophysics and cosmology and supported by BIS National E-infrastructure capital grant ST/J005673/1 and STFC grants ST/H008586/1, ST/K00333X/1. The work of AAK was supported by the Foundation for the Advancement of Theoretical Physics and Mathematics BASIS. The work of A.Y. and J.K. has been supported by the Deutsche Forschungsgemeinschaft (DFG) through the clusters of excellence "Center for Ultrafast Imaging" (CUI, EXC 1074, ID 194651731) and "Advanced Imaging of Matter" (AIM, EXC 2056, ID 390715994).